\documentclass[aps,pra,twocolumn,superscriptaddress,nofootinbib,amsmath,amssymb]{revtex4-2}

\usepackage{lineno}
\usepackage{datetime}

\usepackage{dcolumn}                    
\usepackage{bm}                         
\usepackage[pdftex]{graphicx,color}     
\usepackage{graphicx}                   
\usepackage[caption=false]{subfig}      
\usepackage{float}                      
\usepackage{xcolor}                     
\usepackage{svg}                        
\usepackage[english]{babel}             
\usepackage{url}                        

\begin{document}
%
\title{
Experimentally mapping the scattering phases and amplitudes of a finite object by optical mutual scattering } 
\author{Alfredo Rates} 
\altaffiliation[Present Address: ]{Cell Biology, Neurobiology and Biophysics, Department of Biology, Faculty of Science, Utrecht University, Utrecht, the Netherlands} 
\author{Ad Lagendijk} 
\author{Minh Duy Truong} 
\altaffiliation[Present Address: ]{Hyosung Heavy Industries R\&D Center Netherlands, Arnhem, the Netherlands} 
\author{Willem L. Vos} 
\altaffiliation[Author for correspondence: ]{w.l.vos@utwente.nl} 
\affiliation{Complex Photonic Systems (COPS), MESA+ Institute for Nanotechnology, 
University of Twente, P.O. Box 217, 7500 AE Enschede, The Netherlands}
%
%
\begin{abstract}
Mutual scattering arises when multiple waves intersect within a finite scattering object, resulting in cross-interference between the incident and scattered waves. 
By measuring mutual scattering, we determine the complex-valued scattering amplitude $f$ --- both amplitude and phase --- of the finite object, which holds information on its scattering properties by linking incident and outgoing waves from any arbitrary direction. 
Mutual scattering is present for any coherent wave --- acoustic, electromagnetic, particle --- and we here demonstrate the effect using optical experiments. 
We propose an experimental technique for characterization that utilizes mutual scattering and we present our results for four distinct finite objects: a polystyrene sphere (diameter $59\ \mu$m), a single black human hair (diameter $92\ \mu$m), a strip of pultruded carbon (edge length $140\ \mu$m), and a block of ZnO$_2$ (edge length $64\ \mu$m). 
Our measurements exhibit qualitative agreement with Mie scattering calculations where the model is applicable. 
Deviations from the model indicate the complexity of the objects, both in terms of their geometrical structure and scattering properties. 
Our results offer new insights into mutual scattering and have significant implications for future applications of sample characterization in fields such as metrology, microscopy, and nanofabrication.
\end{abstract}
%
%
\maketitle
%
%
%
\section{Introduction}
%

How waves interact with matter is a topic under extensive study in modern physics, either to study an object using waves as a probe or to study waves using matter for detection. 
This interaction comprises every type of wave, either electromagnetic, particle, or acoustic. 
Scattering is a fundamental effect that occurs when a particle or a wave interacts with a scattering object, or \textit{scatterer}. 
Scattering describes the generation of scattered waves as an incident wave interacts with scatterers. 
The theory of scattering is a mature and unifying branch of theoretical physics~\cite{VanRossum1999rmp, Akkermans2007book, Wiersma2013np}, and scattering measurements of an object is a vast source of information on the properties of the object under study, ever since early 1900s studies of scattering by atomic nuclei~\cite{Goodman1968book}. 
Thus, it is not a surprise to see scattering measurements being used for characterization in a wide range of applications, with the vast range of electromagnetic waves (\textit{e.g.}, dynamic light scattering~\cite{Pecora1985book, Stetefeld2016br}, optical coherent tomography~\cite{Huang1991science, Podoleanu2012jom, Drexler2015book}, transmission-matrix measurements~\cite{Beenakker1997RevModPhys, Kohlgraf2008OptExp, Popoff2010PRL, Akbulut2016PRA}), particle waves (\textit{e.g.}, electron microscopy~\cite{Egerton2016book, spence2013book}, neutron scattering~\cite{boothroyd2020book, willis2017book}), and acoustic waves (\textit{e.g.}, medical ultrasound~\cite{azhari2010book, hoskins2019book}, sonar systems~\cite{kolev2011book, Lurton2016book}). 

\begin{figure}[tbp]
    \centering
    \includegraphics[width=\columnwidth]{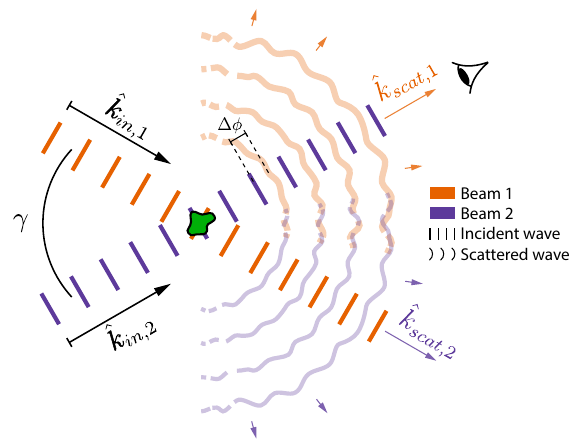}
    \caption{Schematic of two-wave mutual scattering. 
    $N=2$ beams with directions $\hat{k}_{{\rm in},1}$ and $\hat{k}_{{\rm in},2}$ and mutual angle $\gamma$ are incident onto a finite scattering object. 
    The spherical scattered waves are shown as curved wobbly wavefronts to emphasize that they are present at all outgoing directions ($\hat{k}_{{\rm scat},1}$ and $\hat{k}_{{\rm scat},2}$). 
    When $\hat{k}_{{\rm scat},1}=\hat{k}_{{\rm in},2}$ and $\hat{k}_{{\rm scat},2}=\hat{k}_{{\rm in},1}$, the scattered wave interfere with the incident coherent beam. 
    The phase difference between two incident waves is referred to as $\Delta \phi$.
    The colors distinguish the scattered waves and do not represent different wavelengths. 
    }
    \label{fig:currComp}
\end{figure}
%
The scattering amplitude $f$ encapsulates all scattering properties of an object under study, both phases and amplitudes, similar to a transmission matrix, and dictates the relation of an incident wave with the resulting scattered wave at any given angle~\cite{Messiah1967book, newton1982book, lagendijk1996prep, Carminati2021book}. 
Once the scattering amplitude of an object is obtained, all scattering properties can be derived. 
Here, we extract the complex-valued scattering amplitude $f$ by measuring the change of \textit{extinction}. 
Extinction, or to be more precise, \textit{self-extinction}, refers to the attenuation of the incident wave, due to both absorption and scattering, out of the incident wave~\cite{newton1982book}. 
In this paper, we show how to use extinction to characterize any scattering object. 

In many scattering experiments~\cite{Wiersma2013np}, the area of the object is much greater than the area of the incident beam of waves. 
In addition, the object is assumed to be opaque, meaning all incident waves or particles interact with the object. 
This type of measurement induces a relevant yet commonly overlooked limitation: all incident radiation is scattered, and the incident wave is not present at the detector plane in transmission~\cite{ishimaru1978book, Jackson1998book, Akkermans2007book, Carminati2021book}. 

When the object is \textit{finite}, \textit{i.e.}, smaller than the area of the incident radiation, the incident wave is always present at the detector plane when measuring in transmission. 
An advantage of studying finite objects is the ability to observe extinction, which is fully described by interference between the incident wave amplitude, still present in transmission, and the scattered wave amplitude in the forward direction. 
The effect on the incident power of the single incident wave due to extinction is described by the \textit{optical theorem}, a well-known conservation law. 

Recently, our team has discovered that if multiple incident waves hit a finite object, as illustrated in Fig.~\ref{fig:currComp}, there is a cross-interference between the incident coherent wave of one incident beam and the scattered wave of another incident beam~\cite{Lagendijk2020EPL}, an interference effect that is not described by the traditional optical theorem. 
We call this cross interference \textit{cross-extinction} and we generalized the optical theorem to account for both self-extinction and cross-extinction in the case of multiple incident waves. 
Fig.~\ref{fig:currComp} illustrates this process in the case of two incident beams; the incident purple wave in the figure interferes with the scattered orange waves in the direction of the detector. 

Furthermore, we discovered that cross-extinction controls the total extinction of the object. 
The change of the total extinction depends on the angle and phase difference between the incident waves and on the scattering amplitude $f$ of the object. 
Thus, by changing the angle and phase of the incident waves, we tune the extinction of the object, making it appear more translucent (less extinction) or more opaque (more extinction). 
We demonstrated this control of extinction experimentally~\cite{Rates2021PRA}, and we call it \textbf{mutual scattering}. 

In this paper, we present an experimental method and protocol to characterize any scattering object with mutual scattering, unraveling how they interact with light waves. 
By measuring mutual scattering we extract the scattering amplitude $f$. 
To extract $f$ and thus characterize the scattering properties of the object, we need to extract both amplitude and phase information from the mutual scattering measurements. 
We do this by shaping two incident beams with a digital micromirror device (DMD) and measuring the transmitted light. The high stability of a DMD allows for both amplitude and phase extraction. 

To our knowledge, this is the first experiment designed to extract the complex-valued scattering amplitude, including its phase, for every angle. 
Having access to the complex-valued scattering amplitude $f$ is an improvement from conventional intensity measurements, which are proportional to $|f|^2$, and since extinction consists of both absorption and pure scattering components, mutual scattering provides a significant benefit as it enables the characterization of strongly absorbing objects. 
Here, we show the mutual scattering measurements of 4 finite objects with different optical properties: (i) a polystyrene sphere, (ii) a single black human hair, (iii) a strip of pultruded carbon, and (iv) a block of ZnO$_2$. 
For the polystyrene sphere and the human hair, we compare our measurements with Mie calculations.

\begin{figure}[tbp]
    \centering
    \includegraphics[width=\columnwidth]{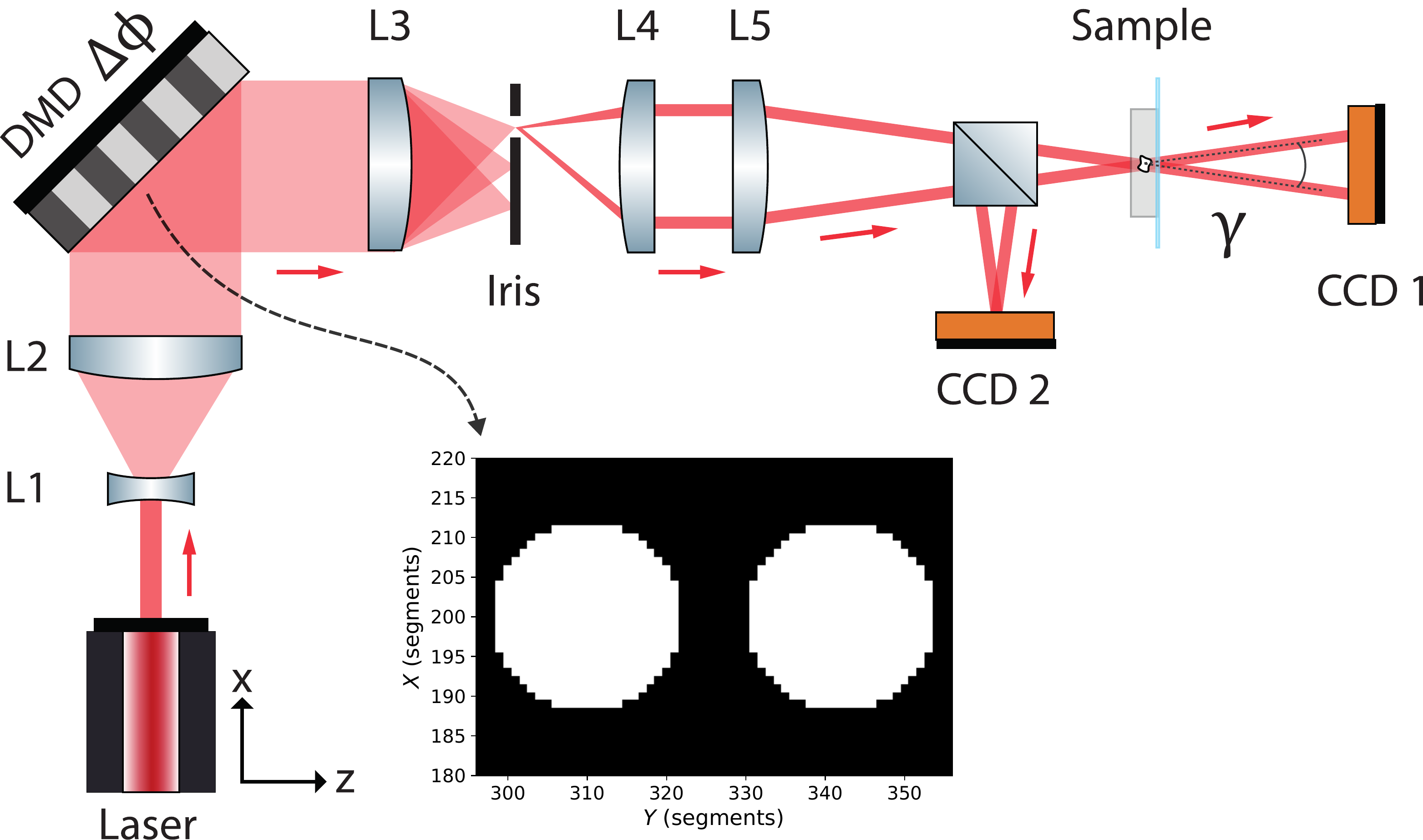}
    \caption{Diagram of the experimental setup. 
    The two incident beams are generated with active areas in the digital micromirror device (DMD), which also sets the phase difference $\Delta \phi$ between the beams. 
    The active areas are shown in the inset. 
    Lenses L3 and L5 compose the holographic filter needed for phase control. 
    The two beams are focused into the sample by lens L5 at an angle $\gamma$, and collected by camera CCD1. 
    A beamsplitter is placed in front of the sample such that the reflected focus arrives on camera CCD2. 
    }
    \label{fig:setupB}
\end{figure}
%
%
\section{Experimental methods}
%
%
\subsection{Optical setup}
%

To study mutual scattering of light, we built the experimental setup shown in Fig.~\ref{fig:setupB}. 
We utilize a He-Ne laser (Hughes 3225H-PC, $5$ mW, $\lambda=632.8\ \rm{nm}$) as a source and a $\times 15$ telescope to enlarge the beam area. 
The angle and phase of the incident beams are controlled by a digital micromirror device (DMD, Vialux VX4100), where the two incident beams are generated with two active areas in the DMD. 
An active area is a collection of adjacent DMD segments (in this case, $24$ segments) that are simultaneously activated, surrounded by inactive segments (see inset Fig.~\ref{fig:setupB}). 

The relative phase $\Delta \phi$ of the beams is controlled using the Lee holography technique~\cite{Lee1978progrInOpt, Conkey2012optexp}. 
We implement the holographic filter needed for this technique with lenses L3 and L4 and an iris in between. 
The two beams are then focused into the sample with lens L5 in such a way that the beams form an angle $\gamma$. 
To change the angle $\gamma$, we change the position of the active areas of the DMD, thus making the beams closer or further apart before reaching lens L5 (see inset Fig.~\ref{fig:setupB}). 
Two charge-coupled device cameras (CCD, Stingray F-125) detect the flux $F$ integrated over the illuminated pixels. 
The flux $F$ measures the energy flow in the detector area and is the magnitude of the Poynting vector $S$, unlike the intensity $I$, which measures power per area. 
Camera CCD1 measures the flux of the two incident beams transmitted along the sample, whereas camera CCD2 is placed at the focal distance of L5, and its measurement corresponds to the wavefront pattern at the sample position. 
We use camera CCD1 to measure mutual scattering, while camera CCD2 is used for calibration and characterization. 

\begin{figure}[tbp]
    \centering
    \includegraphics[width=\columnwidth]{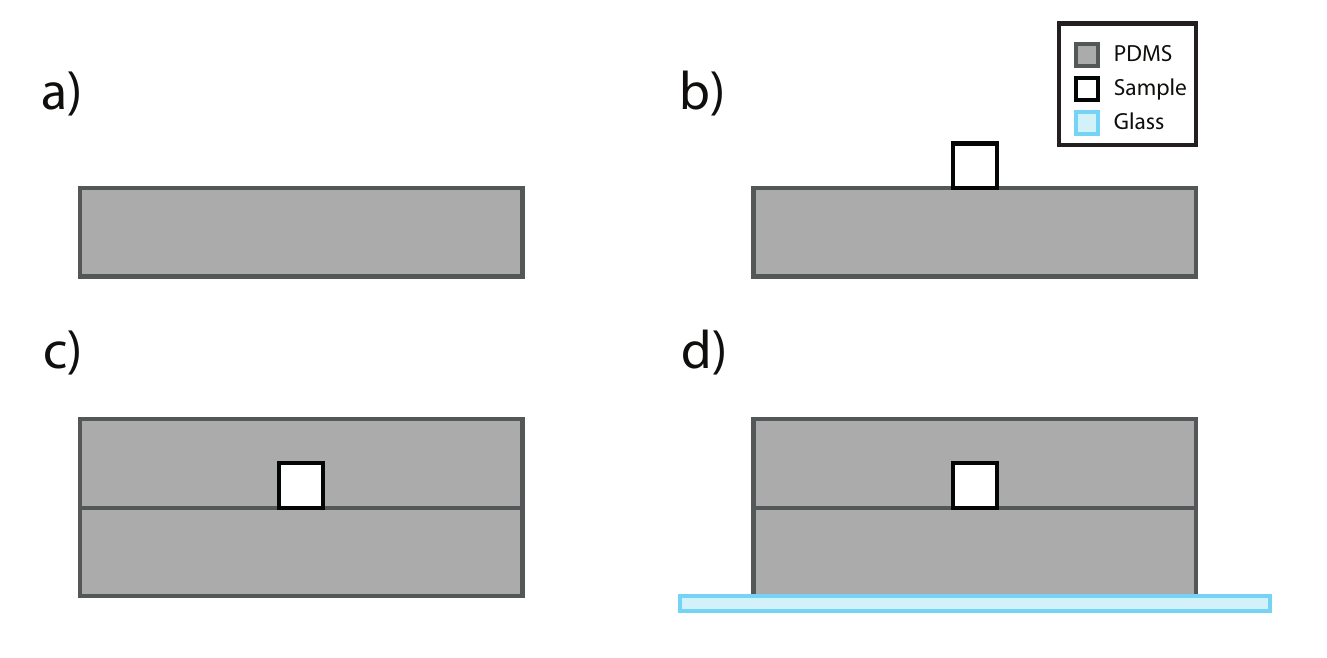}
    \caption{
    Illustration of the main steps in the sample preparation. 
    a) In a Petri dish, we prepare a layer of PDMS (grey). 
    b) Once the PDMS is cured, we place the sample (white) on top of it. 
    c) We cover the sample with a new layer of PDMS. 
    d) Once the second layer is cured, we cut out the relevant part of the sample and glue it on a microscope glass slide. 
    }
    \label{fig:samplePrep}
\end{figure}
%
The measurement consists of a total of 280 positions of $\gamma$ from $0^\circ$ to $2.8^\circ$. 
For each angle $\gamma$, we vary the phase difference $\Delta \phi$ from $0$ to $2\pi$ with 30 steps in between. 
To extract the normalized mutual scattering component $F_{{\rm MS}}$ we follow the same procedure described in our previous work~\cite{Rates2021PRA}, as
%
\begin{align}\label{eq:totalExtinction}
    F_{{\rm MS}}  =    F^{\rm ws}_{1,2} - F^{\rm ws}_{1} - F^{\rm ws}_{2} - \left( F^{\rm ns}_{1,2} - F^{\rm ns}_{1} - F^{\rm ns}_{2} \right),
\end{align}
%

where the sub-index labels which beam is activated (only beam 1, only beam 2, or both 1,2), and the upper-index labels whether we measure with the sample present (ws) or when there is no sample present (ns). 
The component $F_{{\rm MS}}$ represents the change of the extinction flux of light due to mutual scattering and it is normalized with the total extinction of a single beam. 
If $F_{{\rm MS}}=0$, the extinction has the same value as in a single-beam experiment; if $F_{{\rm MS}}=1$, the extinction is twice as large, making the object twice as opaque; and if $F_{{\rm MS}}=-1$, the extinction is canceled out, making the object fully transparent.

\begin{figure}[tbp]
    \centering
    \includegraphics[width=\columnwidth]{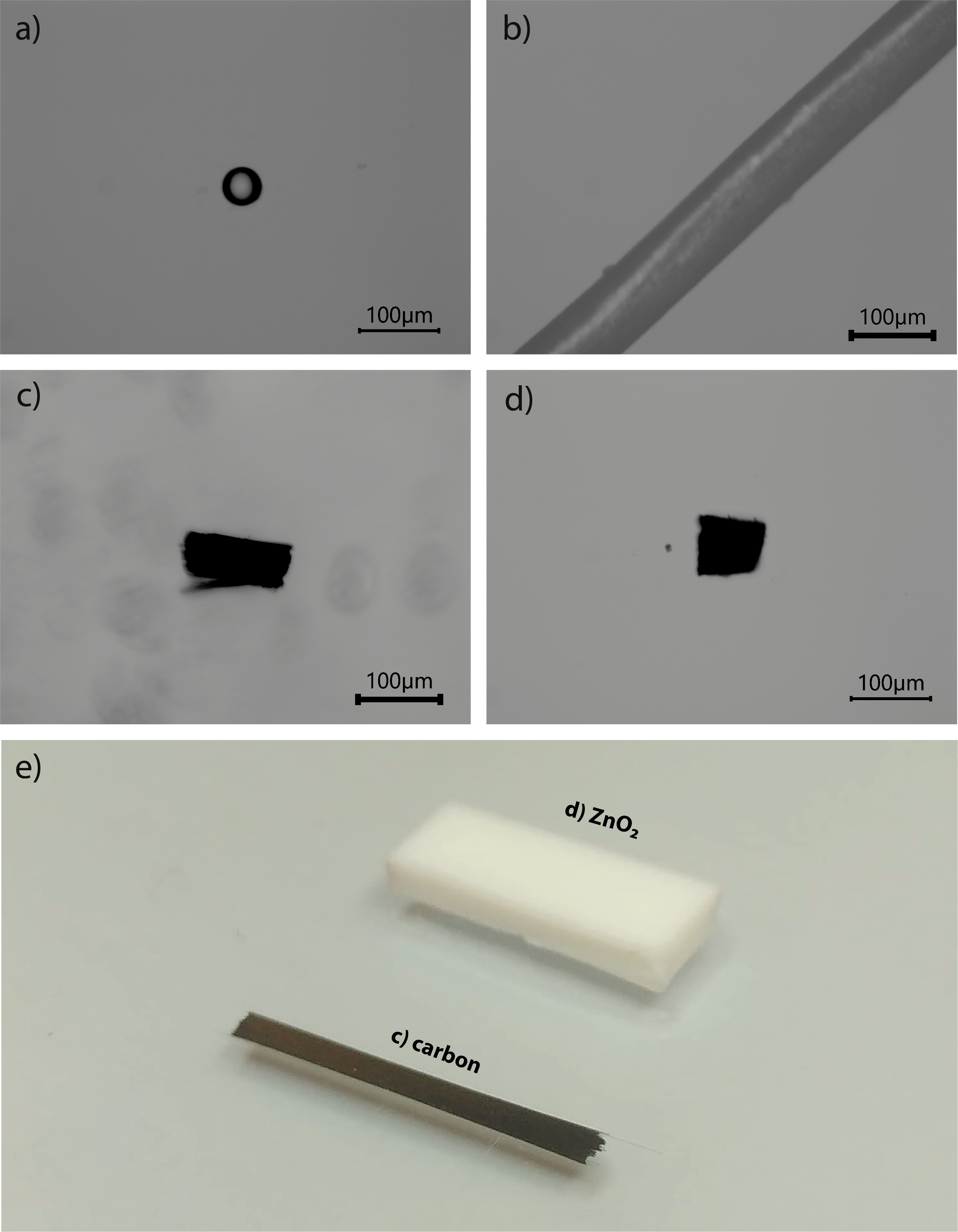}
    \caption{
    Microscope images of the samples studied. 
    a) Polystyrene sphere, b) a single black human hair, c) a pultruded carbon strip, and d) a block of scattering ZnO$_2$ nanospheres. 
    All pictures are taken at the end of the fabrication process, so all samples except the hair are embedded in two stacked layers of PDMS. 
    e) Image of macroscopic raw materials, namely pultruded carbon and ZnO$_2$ scattering sample. 
    Both samples were manually cut to the desired geometry. 
    }
    \label{fig:samples}
\end{figure}
%
\begin{figure}[tbp]
    \centering
    \includegraphics[width=\columnwidth]{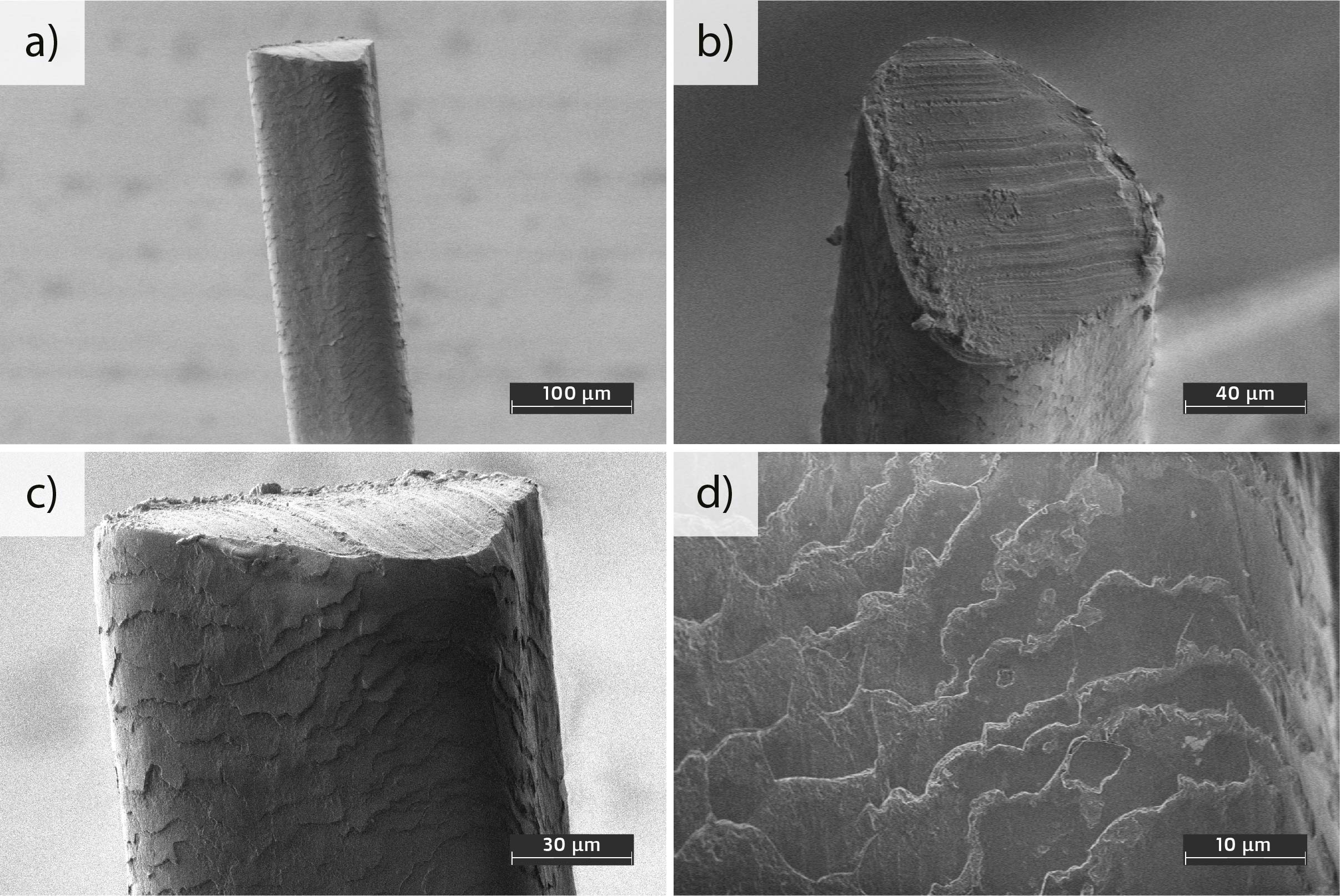}
    \caption{SEM images of the human hair sample for different positions and magnifications. 
    }
    \label{fig:semHair}
\end{figure}
%
\newpage
%
\subsection{Samples}
%

\begin{table}[tbp]
    \centering
    \begin{tabular}{p{0.4\columnwidth}|p{0.2\columnwidth}|p{0.15\columnwidth}|p{0.15\columnwidth}}
    Sample             & $n_{\rm s}$ & $n_{\rm m}$ & $2r_{\rm s}$ ($\mu$m) \\ \hline
    Polystyrene sphere & 1.5875$^1$      & 1.412$^1$       & 59                    \\
    single black hair   & 1.55+1i~\cite{Marschner2003acm}     & 1           & 92                   
    \end{tabular}
    \caption{Parameters used for our Mie calculations. 
    \\
    ($^1$From manufacturer)}
    \label{tab:paraModel}
\end{table}
%
We measure $F_{{\rm MS}}$ for four different samples: (i) a polystyrene sphere (Thermo Scientific 7000 Series, 7550A $55\ \mu$m), (ii) a single black human hair, (iii) a strip of pultruded carbon, and (iv) a block of ZnO$_2$. 
With the exception of the single human hair, all samples are embedded in PDMS (Dowsil SYLGARD 184 \& included curing agent). 
The samples are prepared with two layers of PDMS, as shown in Fig.~\ref{fig:samplePrep}. 
First, we prepare a PDMS layer in a Petri dish and employ a vacuum pump to remove the microbubbles in the PDMS. 
Once the PDMS is cured, we place the sample on top of it and fill a second layer following the same procedure. 
When the second layer is ready, we cut the area where the sample is embedded and glue it to a microscope glass slide. 
For all steps, the proportion between PDMS and the curing agent is 10:1. 
Meanwhile, we glue both ends of the single hair to a microscope glass slide such that it is suspended in air. 

A microscope picture of every sample already embedded in PDMS is presented in Fig.~\ref{fig:samples}. 
Note that due to a microscope artifact, the ZnO$_2$ in Fig.~\ref{fig:samples}d looks as opaque as the pultruded carbon in Fig.~\ref{fig:samples}c. 
To better showcase the difference in optical properties, Fig.~\ref{fig:samples}e shows a picture of the bulk materials. 
In addition, Fig.\ref{fig:semHair} shows scanning electron microscope (SEM) images of the surface and cross-section of the human hair. 
The refractive index $n$ of the sphere and the hair are shown in Table~\ref{tab:paraModel}. 

While the sphere and the hair already have the desired dimensions ($2r_{\rm sphere} = 59\ \mu$m and $2r_{\rm hair} = 92 \ \mu$m), the pultruded carbon and the ZnO$_2$ need to be manually prepared. 
The pultruded carbon is prepared from a commercially available manufacturer (van Dijk Pultrusion Products DPP) with dimensions $1000\times10\times0.1$ mm, and it is manually cut to a cuboid with estimated dimensions $140\times54\times10 \ \mu$m. 

The block of ZnO$_2$ is prepared using ZnO$_2$ particles (Sigma-Aldrich 544906 Zinc Oxide nanopowder) suspended in PDMS, following the same procedure discussed above, with a particle concentration of 10\%. 
Then, the sample is manually cut into a cuboid with an estimated edge length of $L_{\rm ZnO_2} = 64 \ \mu$m. 

We position all samples at the crossing point of the two incident beams, \textit{i.e.}, at the focal distance of L5. All the samples have a size smaller than the beam waist at the focus. 
Thus, they do not block all the light, and the incident beam is still present at the detector.

%
\section{Experimental results}
%

%
\subsection{Wavefront correction}
%

A crucial aspect of the experiment is the calibration of the incident beams. 
Ideally, when using a plano-convex lens, all the rays cross at the focal point. 
Hence, two incident beams are sure to cross on the sample. 
Unfortunately, any misalignment or aberration will distort the focus. 
In addition, we realized the DMD itself causes strong aberrations in the wavefront. 
This is most probably because the micro-mirrors are not perfectly flat, and any curvature or imperfection adds up to the final wavefront. 

To overcome this issue, we implement the correction shown in Ref.~\cite{Popoff2019website}. 
The correction is based on optimizing the intensity pattern at the focus by adding a phase mask and a pattern generated by Zernike polynomials. 
This correction is based on the assumption that common aberrations have a pattern described by Zernike polynomials~\cite{Born1970book}.
In Ref.~\cite{Popoff2019website}, they iterate the coefficients of the polynomials to maximize the intensity at the focal point. 
For our case, we found that instead of maximizing intensity, we obtain a better correction by calculating beforehand the expected intensity distribution at the focus and then maximizing the 2D correlation between the measurements and the calculated pattern. 
For this calibration, we employ the camera CCD2 (see Fig.~\ref{fig:setupB}). 
Having a beamsplitter adds the possibility of calibrating the setup without disturbing the sample. 

\begin{figure}[tbp]
    \centering
    \includegraphics[width=\columnwidth]{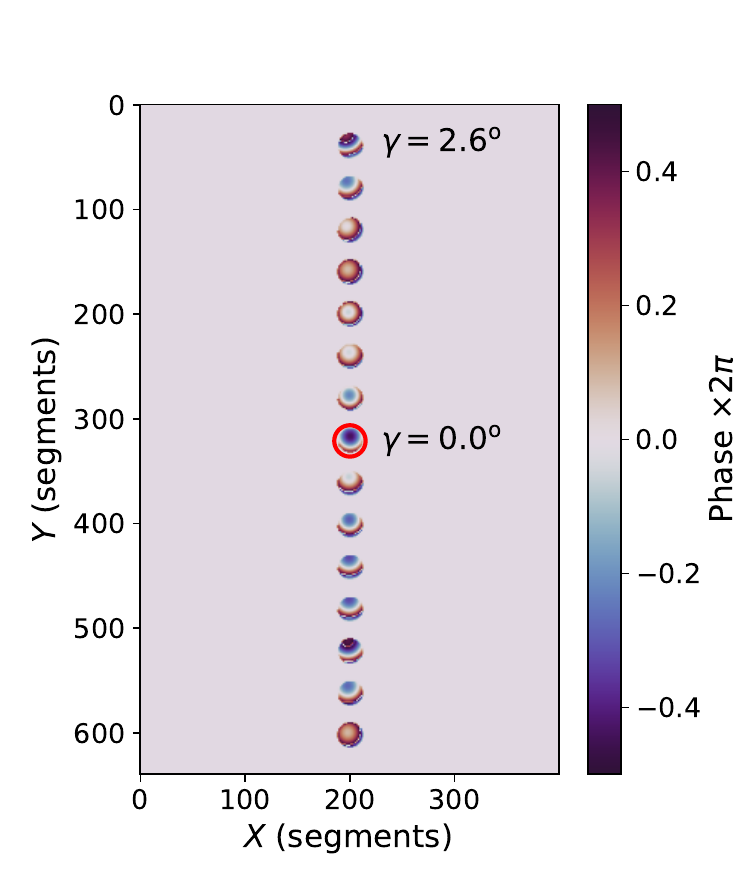}
    \caption{Phase masks in DMD for selected angles. 
    Each circle corresponds to a different incident angle at the sample. 
    The red circle highlights the central circle, which is always activated. 
    All the segments outside the circles have phase \textit{and amplitude} 0. 
    The color bar is circular (the color at $1$ is the same as the color at $-1$) to account for the periodicity of the phase. 
    }
    \label{fig:dmdPatt}
\end{figure}

In Fig.~\ref{fig:dmdPatt}, we show an example of the phase masks used for different angles. 
We characterize our experiment by running the optimization described above for every active area separately. 
In the experiments, only two areas are simultaneously activated for a given angle. 
For the results shown in this paper, we keep the center area (highlighted with a red circle) constantly activated, and only vary the second beam. 
We call this the \textit{asymmetric case}. 
Still, our experimental setup also allows for the \textit{symmetric case}, where we vary both beams simultaneously. 
In our system, it is also possible to vary the position of the active areas not only in one line but at any angle by characterizing the whole DMD. 
Note that here we only show a set of selected angles, while in reality, we have almost 600 different positions along the y-axis of the DMD.

Once a pair of areas is activated and the angle $\gamma$ is set, we scan the phase from $0$ to $2\pi$ with 30 steps in between. 
We iterate this scan three times to have better statistics. 
We extract the amplitude and phase of mutual scattering by taking the Fourier transform of the phase-dependent curve and filtering the frequency $f=1/2\pi$. 
We then take the maximum value, the minimum value, and the phase of the curve. 
Examples of the phase-dependent curve are shown in Fig.~\ref{fig:PhaseIter} for angles $\gamma=0.8^\circ$, $\gamma=1.1^\circ$, and $\gamma=2.7^\circ$, from the measurements of the dielectric sphere. 
While most data points have a well-defined shape such as in Fig.~\ref{fig:PhaseIter}a), an example of an undefined phase can be seen in Fig.~\ref{fig:PhaseIter}b), where the extinction amplitude is so small that a single frequency does not describe the data. 
Fig.~\ref{fig:PhaseIter}c) shows an example of a large angle, where the phase has higher uncertainty but it is not yet undefined.

%
%
\subsection{General observations}
%

\begin{figure}[tbp]
    \centering
    \includegraphics[width=\columnwidth]{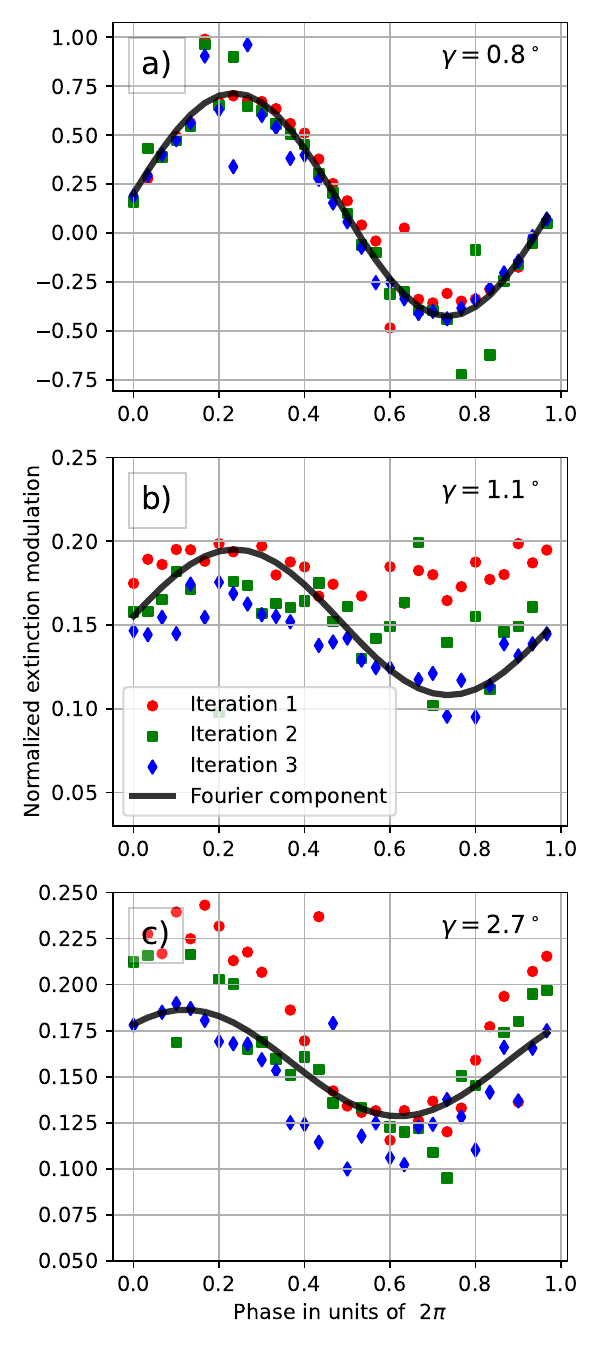}
    \caption{Total extinction flux against phase difference between beams for angles a) $\gamma=0.8^\circ$, b) $\gamma=1.1^\circ$, and c) $\gamma=2.7^\circ$. 
    Circle symbols are experimental data, and the different colors represent the different iterations of the measurements. 
    The solid black line shows the extracted cosine component with Fourier filtering. 
    }
    \label{fig:PhaseIter}
\end{figure}

\begin{figure*}[tbp]
    \centering
    \includegraphics[width=\textwidth]{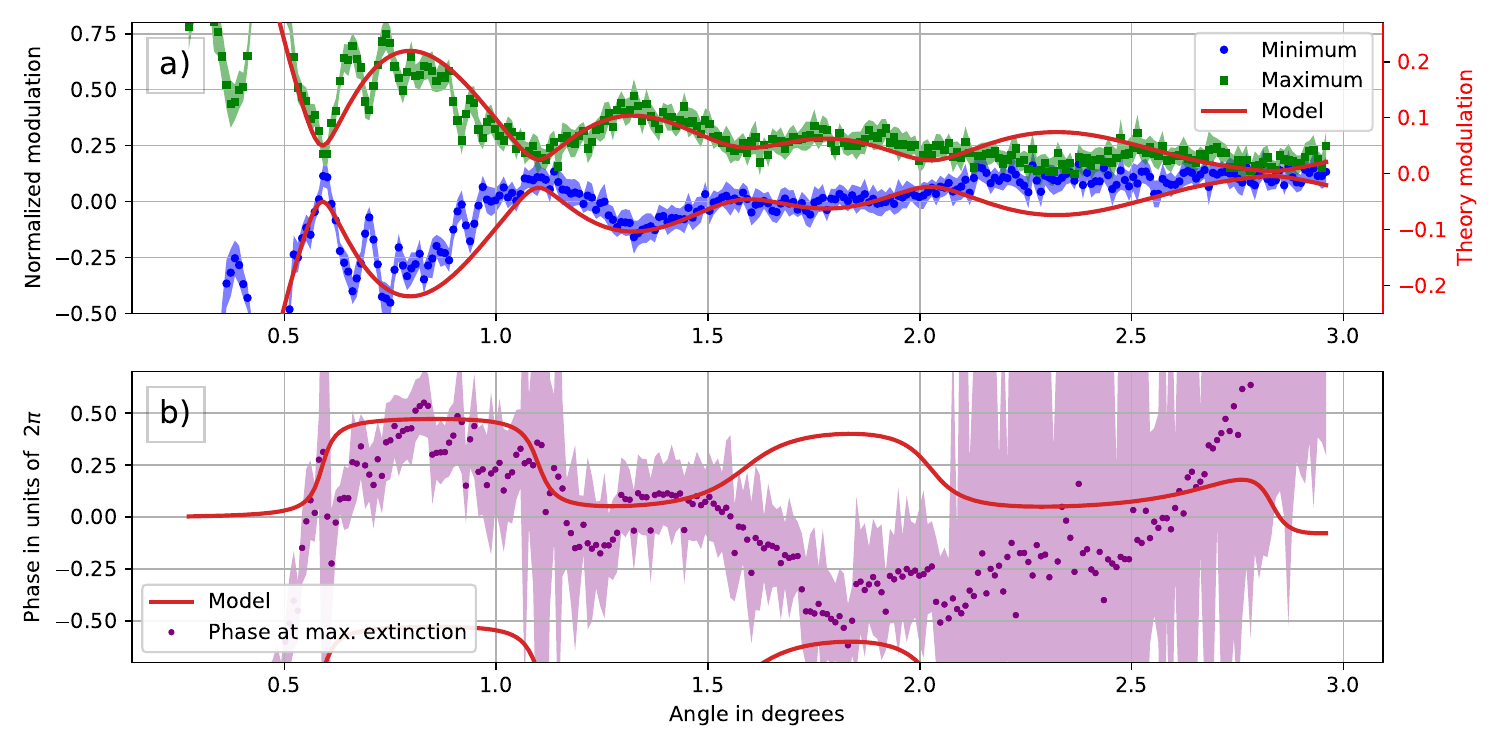}
    \caption{
    Mutual scattering of the $59-\mu$m polystyrene sphere. 
    (a) Amplitude of the total extinction change, with green squares the measured maximum extinction, blue circles the minimum extinction, and the red curve our model (plotted versus right red y-axis). 
    (b) Phase of the total extinction change, with purple circles the measured phase and the red curve our model. 
    }
    \label{fig:resSphere}
\end{figure*}

\begin{figure*}[tbp]
    \centering
    \includegraphics[width=\textwidth]{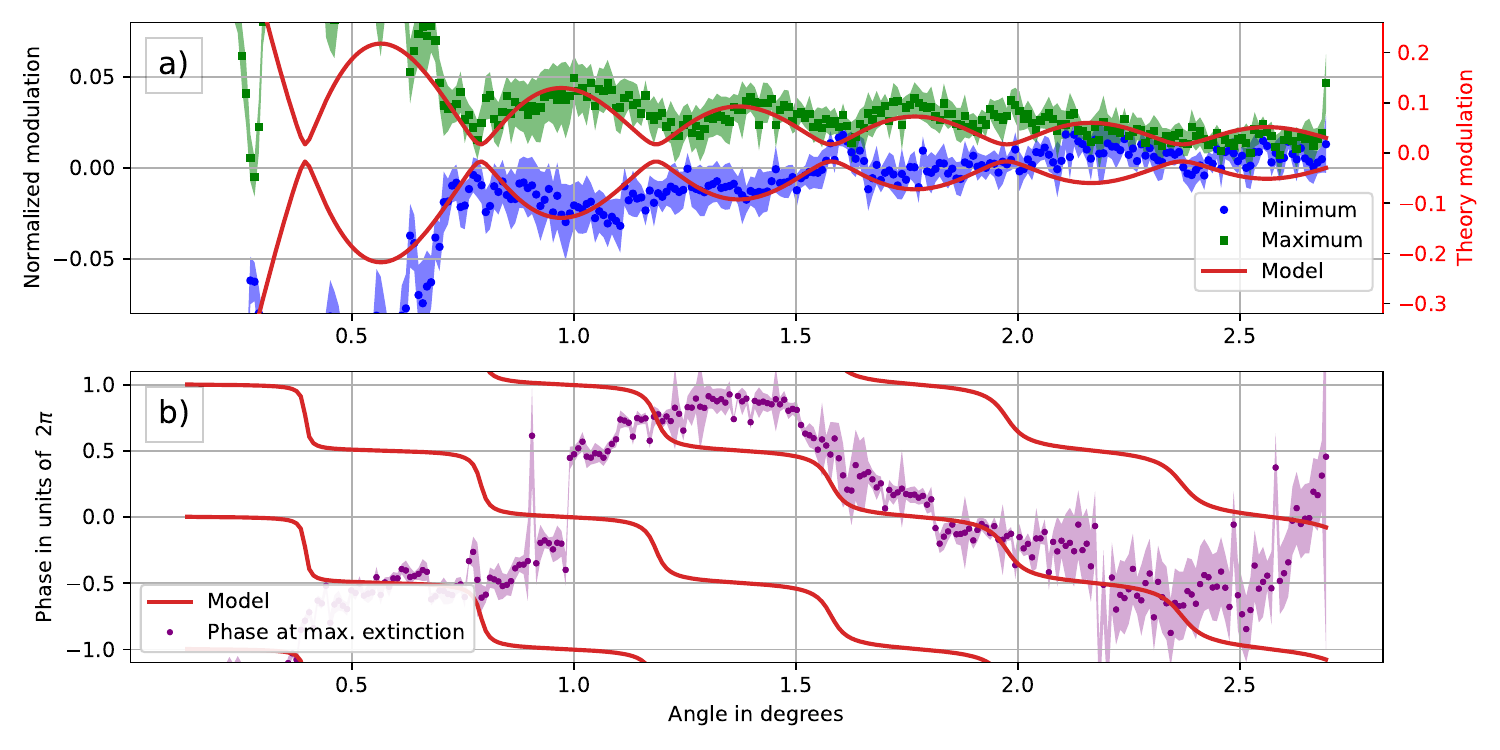}
    \caption{
    Mutual scattering result for single black human hair. 
    (a) Amplitude of the total extinction change, with green squares the measured maximum extinction, blue circles the minimum extinction, and the red curve our model (plotted versus right red y-axis). 
    (b) Phase of the total extinction change, with purple circles the measured phase and the red curve our model. 
    }
    \label{fig:resHair}
\end{figure*}

Four samples were measured: a dielectric sphere, a single black human hair, a strip of pultruded carbon, and a block of ZnO$_2$. 
The results for all the samples are present in Figures~\ref{fig:resSphere} to~\ref{fig:resZnO2}. 
The amplitude of mutual scattering is represented by the total modulation of the extinction, shown in the top panel of each figure, where the blue circles and green squares represent the minimum and maximum extinction, respectively. 
These values are extracted from the phase dependency at each angle, as shown in Fig.~\ref{fig:PhaseIter}.

%

%
Additionally, the solid red lines in Fig.~\ref{fig:resSphere} and Fig.~\ref{fig:resHair} show the results from our Mie scattering calculations, both for amplitude and phase. 
Note that we plot the model against the right y-axis, which has a scaling factor and offset for clear comparison. 
We include multiple model curves with $2\pi$ spacing to account for the periodicity of the phase. 
For our calculations, we assume a sphere geometry for the dielectric sphere and a cylinder geometry for the single human hair. 
The input to the model is the refractive index of the sample $n_{\rm s}$, the refractive index of the medium $n_{\rm m}$, and the sample radius $r_{\rm s}$. 
The parameters used for the models are presented in Table~\ref{tab:paraModel}. 
These Mie models are not applicable to the strip of pultruded carbon and the block of ZnO$_2$ due to their complicated shape. 

The bottom panel of each sub-figure of Figures~\ref{fig:resSphere} to~\ref{fig:resZnO2} shows the angular dependency of the phase information, which is the phase at which we obtain the maximum value (green squares in the top panel). 
The phase information of mutual scattering allows the extraction of the complex-valued scattering amplitude $f$ and, to our knowledge, this is the first experiment where the phase information is available for any given angle. 
The calibration of the phase information is discussed in Appendix~\ref{sec:appA}. 
When the amplitude of mutual scattering is close to zero, the phase has an increasing error margin. 
We consider these points as \textit{undefined} (see Fig,~\ref{fig:PhaseIter}b)). 

%
\subsection{Polystyrene sphere}
%

In Fig.~\ref{fig:resSphere}, we see that the measurements of the polystyrene sphere have a trend that is in agreement with our model, both in amplitude and phase. 
In a previous work~\cite{Rates2021PRA}, we show how analytical models are in agreement with experiments when the angle is close to zero $\gamma \to 0$, and deviate when increasing the angle, as we approach the \textit{speckle regime}, where the distribution of scatterers and the exact 3D shape of the object gain more relevance. 
Here we observe a similar behavior, with the model for the amplitude information starting to deviate around $\gamma=2.1^\circ$. 
Around $\gamma=1.5^\circ$, the model and the data seemingly split; the model goes upwards while the data goes downwards. 
Although the trend is different, the phase goes down to $-0.5\times 2\pi$, which is close to the value of the model after accounting for the periodicity of the phase. 

The measurements show that the phase becomes undefined at the \textit{nodes} of the amplitude, namely around $\gamma=0.6^\circ$ and $\gamma=1.2^\circ$. 
We call these \textit{closed nodes}, where the control range of the extinction is so small that it is not possible to define the optimal phase. 
Still, a third node is present around $\gamma=1.6^\circ$ which is not accompanied by an undefined phase. 
We call these types of nodes \textit{open nodes}, as the gap between maximum and minimum extinction is not completely closed. 

%
The measurements of the polystyrene sphere show a larger error range compared to the other samples. 
This is mainly due to the low contrast of the refractive index between the sample and the surrounding media (see Table~\ref{tab:paraModel}). 
This translates into a weak scattering signal and, thus, a lower signal-to-noise ratio (SNR). 
This is particularly relevant at the end of the angular range for the amplitude data, where the low SNR makes the phase undetermined. 
Interestingly, the phase data starts deviating from the theory just at the point of the expected third node, which is an open node, located before the amplitude deviates from the theory. 

A possible reason for the differences between theory and experiments is the geometry of the sample. 
While the Mie calculations assume a perfect, homogeneous sphere, our sphere is exposed to fabrication imperfections. 
For example, it has been previously observed~\cite{Megens1997langmuir} that a dielectric sphere fabricated by a two-step growing process has the scattering properties of two spheres, one inside the other. 
Our experiment is particularly sensitive to such effects when increasing the angle $\gamma$, as we are entering the speckle regime. 

%
\subsection{Human hair}
%

For the measurements of the single human hair shown in Fig.~\ref{fig:resHair}, we observe that the model is in good agreement with the experimental measurement of the amplitude of mutual scattering. 
Similar to the polystyrene sphere, when increasing the angle we approach the speckle regime and the model deviates from the experimental data. 
The separation happens around $\gamma=1.9^\circ$, although for small angles we also note a difference between the model and the data, as the first node of the model is not present in the experimental data. 
Note that for very small angles, the two incident Gaussian beams overlap in the area of the detector, and thus the measurements are less precise. 

For the phase measurements, we see an abrupt jump of $\delta\phi\approx\pi$ around $\gamma=1^\circ$, which is not accounted for in the model. 
After this, the phase information follows the same decreasing trend as the analytical model. 
We also notice that even though we have nodes in the amplitude measurements, the phase is never undefined, even when $\gamma>1.9^\circ$ where the model starts deviating from the amplitude data. 

In this case, the Mie calculation is useful as a first approach, but incomplete. 
We can apply it to, \textit{e.g.}, extract the width of the sample, which dictates the periodicity of the nodes in the amplitude data; we note that the periodicity of the nodes in the sphere measurements is larger than the periodicity in the hair measurements, meaning the sphere is smaller than the hair. 
Nevertheless, the complexity of a biological sample as a single human hair is not encapsulated in our calculations, where we assume the sample to be a perfect cylinder. 

It is well known that human hair is composed of layers, namely the cortex, medulla, and cuticle~\cite{Rogers2019cosm, Yang2014yf}. 
Each layer has different scattering properties, and it may not be homogeneous along the hair~\cite{Marschner2003acm, KharinSPIE2009}. 
Furthermore, the surface of the human hair may contain overlapping cuticle cells that form a scale-like structure, deviating even more from the ideal structure used in the calculation. 
The SEM shown in Fig.\ref{fig:semHair} provides an insight into the structure of human hair, revealing deviations from the assumed cylindrical shape. 
The cross-sectional images presented in Fig.\ref{fig:semHair}a),b) show that the single hair in our sample seems to have a triangular shape. 
Additionally, Fig.~\ref{fig:semHair}c),d) highlight the presence of scale-like structures, which as we mentioned before, are not accounted for in the Mie calculation. 

\begin{figure*}[tbp]
    \centering
    \includegraphics[width=\textwidth]{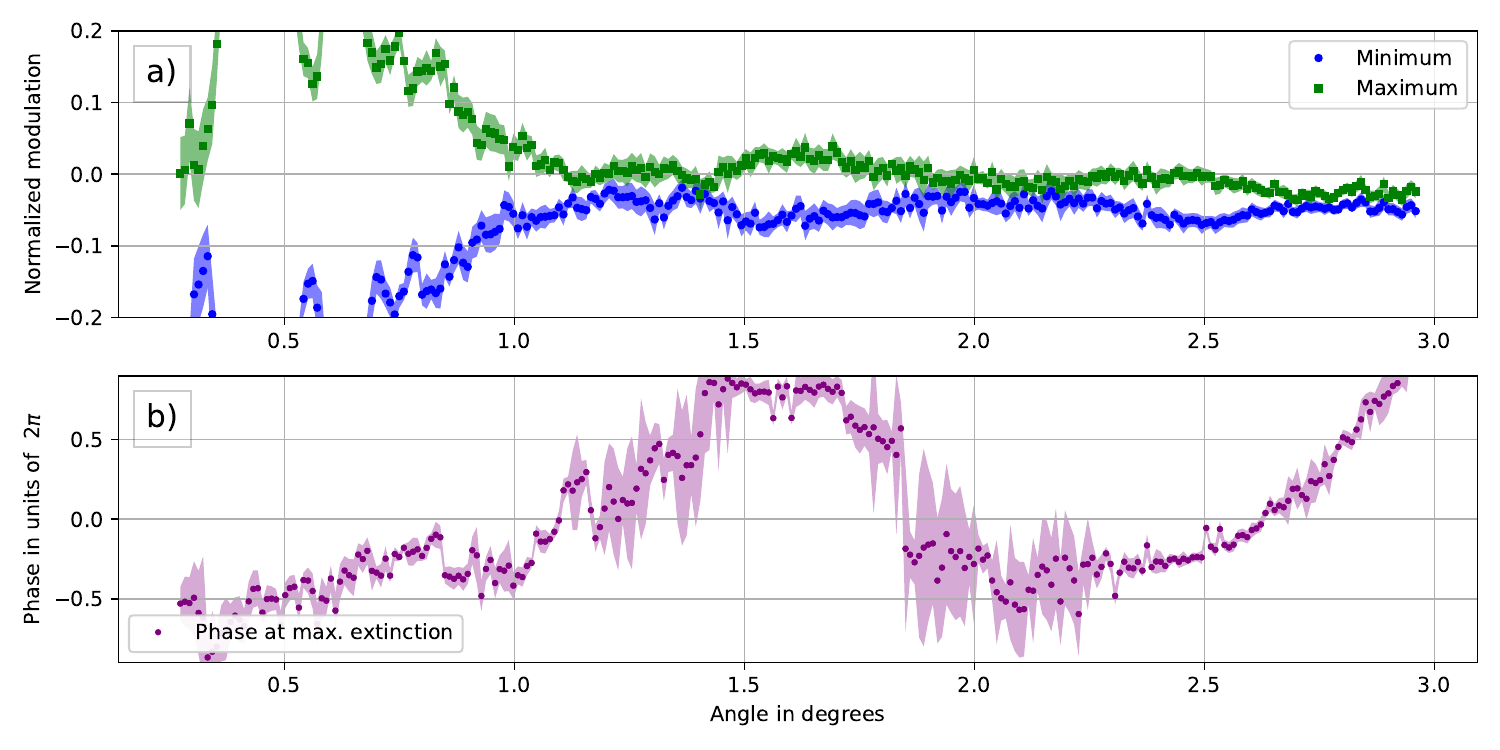}
    \caption{
    Mutual scattering result for carbon strip. 
    (a) Amplitude of the total extinction change, with green squares the measured maximum extinction and blue circles the minimum extinction. 
    (b) Phase of the total extinction change, with purple circles the measured phase.
    }
    \label{fig:resCarbon}
\end{figure*}
%
In addition, the calculation imposes a certain orientation of the sample with respect to the incident beam. 
Thus, any tilt in the sample is reflected in the measurements. 
This is not a problem in the case of the sphere because of its symmetry.

%
%
\subsection{Pultruded carbon strip}

For the amplitude data of the pultruded carbon in Fig.~\ref{fig:resCarbon}, we identify three nodes, namely around $\gamma=1.25^\circ$, around $\gamma=2.1^\circ$, and around $\gamma=2.8^\circ$. 
These nodes are not as clear as for the rest of the samples, and other interpretations are possible, \textit{e.g.}, instead of a single node around $\gamma=1.25^\circ$, there might be two nodes close to each other, around $\gamma=1.2^\circ$ and $\gamma=1.45^\circ$. 
For the case of the phase data, we note a long step between $\gamma=1.2^\circ$ and $\gamma=1.8^\circ$, where the phase is also undefined. 
These positions are close to the nodes described above. 

Due to the free-form shape of this sample, it is very challenging to build an analytical mathematical model to predict experimental results, which is the type of challenge we address in our research program ``Free form scattering optics (FFSO)''. 
We may, in turn, use a simpler approximate model based on diffraction theory, with which we model the sample as a perfectly flat, fully absorbing object. 
One of the many problems with this approach is that it is highly sensitive to the geometry and orientation of the sample. 
Even supposing the sample has a perfect cuboid shape, mutual scattering depends heavily on the orientation of the sample with respect to the axis where the angle $\gamma$ between the two incident beams is formed. 
We hypothesize that the orientation of the carbon strip is the reason why the first node is not so clear for this measurement. 
%
%
\subsection{Block of ZnO$_2$}
%

\begin{figure*}[tbp]
    \centering
    \includegraphics[width=\textwidth]{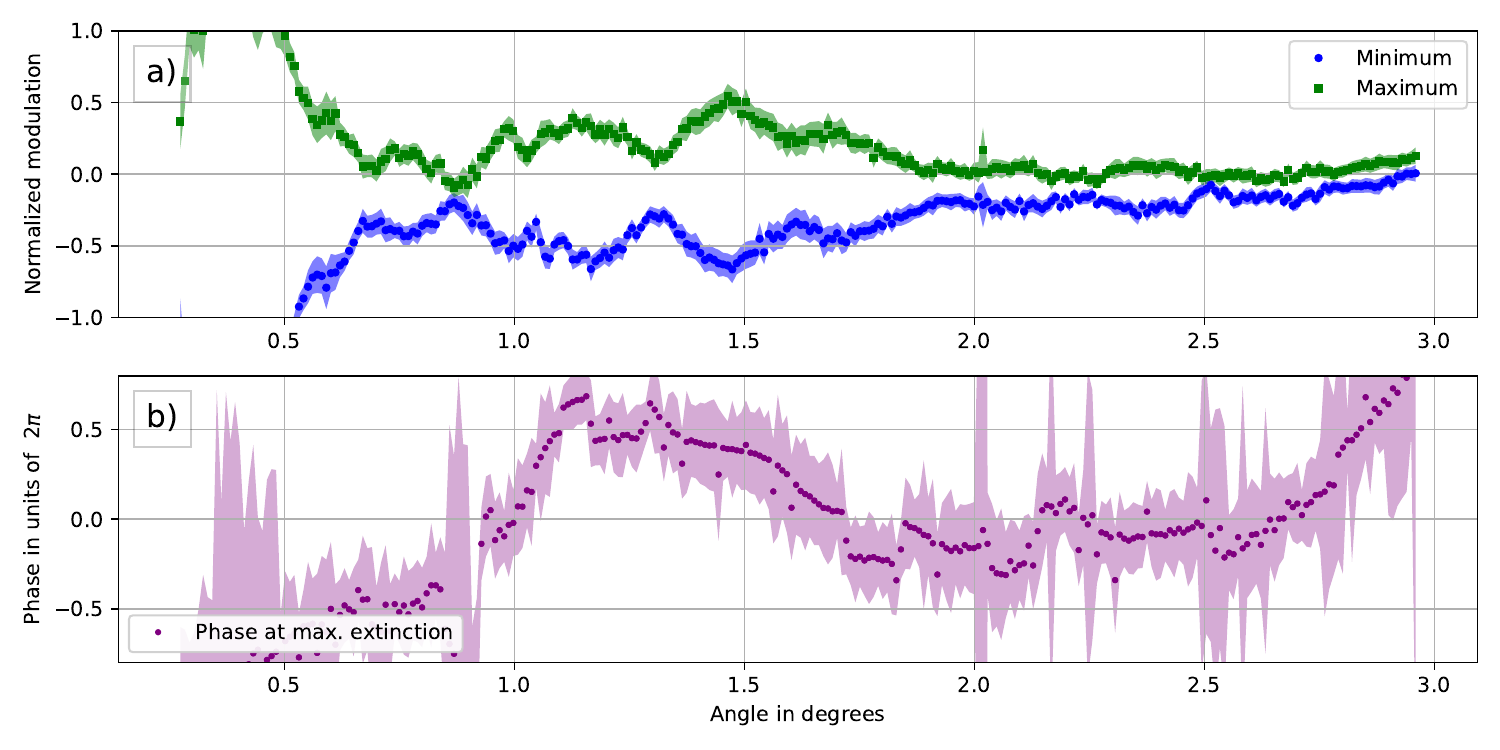}
    \caption{
    Mutual scattering result for ZnO$_2$ block. 
    (a) Amplitude of the total extinction change, with green squares the measured maximum extinction and blue circles the minimum extinction. 
    (b) Phase of the total extinction change, with purple circles the measured phase.
    }
    \label{fig:resZnO2}
\end{figure*}

In Fig.~\ref{fig:resZnO2}, we identify many nodes in the amplitude measurements at specific angles. 
These nodes are located at $\gamma=0.69^\circ$, $0.87^\circ$, $1.03^\circ$, $1.3^\circ$, $1.62^\circ$, $2.0^\circ$, $2.17^\circ$, $2.27^\circ$, and $2.5^\circ$. 
Some of these nodes also have undefined phases. However, there are four nodes at $\gamma=0.69^\circ$, $1.03^\circ$, $1.3^\circ$, and $1.62^\circ$ that do not have undefined phase, which we call open nodes. 
As in the previous cases, the phase data shows a long step between the node at $\gamma=0.87^\circ$ and the node at $\gamma=2.0^\circ$. 
This step seems to agree with an overall increase in the amplitude data. 

%
To better analyze the amplitude data, we divide it into two functions: 
One with a strong angle dependency, known as the carrier signal, and another with a weak angle dependency, known as the envelope signal. 
The carrier signal is responsible for the nodes that are present across the entire angular range and is in sync with the points of undefined phases. 
The envelope signal, on the other hand, is responsible for the overall increase in the amplitude data and is in sync with the step present in the phase data. 
We attribute these two signals to different characteristics of the free-form nature of the sample. 
The carrier signal is related to the object's overall shape, while the envelope signal is related to small changes in geometry. 
Furthermore, the small jumps in phase, such as those observed at $\gamma=1.17^\circ$, $1.85^\circ$, and $2.14^\circ$, are indicative of the speckle behavior of mutual scattering.
%
%
\section{Conclusions}
%

In this work, we present an experimental procedure to measure the change of light extinction in different materials. 
This is done with two incident laser beams and exploiting mutual scattering. 
By varying the relative angle and phase between the two beams, we obtain the control range of the extinction and the phase at which the extinction is maximized. 
From this measurement, it is possible to extract the complex-valued scattering amplitude, including phase, which fully describes the light-scattering properties of the object (see Appendix~\ref{sec:appB}). 
This work shows, to our knowledge, the first mapping of the phase information of the scattering amplitude for any arbitrary angle, by the innovative use of a spatial DMD modulator. 

We studied four different finite scattering objects: a dielectric sphere, a single black human hair, a strip of pultruded carbon, and a block of ZnO$_2$. 
With all these samples, we cover objects with different geometries, objects that are either absorbing or scattering, and we also include a biological sample. 
We compared the polystyrene sphere and the single human hair to Mie calculations, and we found that, especially for the amplitude data, the model is in close agreement with the data. 
In turn, deviations from the model are attributed to object characteristics that differ from the ideal, homogeneous geometries considered in the calculations. 

One of the advantages of our approach with mutual scattering is that it is applicable for either very absorbing or very scattering materials. 
Furthermore, as long as the approximation of far-field holds, mutual scattering is applicable for any object size and wavelength.  
By slightly modifying the setup with the right phase modulators, we can adapt the experiment to have a better angular resolution or to extend the angular range. 
On top of that, a DMD allows us to rotate the plane of the two incident beams or even add more incident beams, which resembles the multiple-beam wavefront shaping setup~\cite{Vellekoop2007OptLett, Vellekoop2008PRL}. 

Finally, mutual scattering is applicable to any type of wave, and we believe that extracting the complex-valued scattering properties of an object has applications beyond optics. 
In particular, mutual scattering has a high potential for application in acoustic and radio waves. 

%
%
\begin{acknowledgments}
%

We thank Cornelis Harteveld for expert technical support and sample preparation, Melissa Goodwin for sample imaging, and Catalina Garc\'ia for her contribution. 
We thank Lars Corbijn van Willenswaard and Bert Mulder, as well as Arie den Boef, Teus Tukker, and Ferry Zijp (ASML) for fruitful discussions. 
This work was supported by the NWO-TTW program P15-36 ‘Free-form scattering optics’ (FFSO) in collaboration with TUE and TUD and with industrial partners ASML, Demcon, Lumileds, Schott, Signify, and TNO, and by the MESA+ Institute's section Applied Nanophotonics (ANP). 
The data used for this publication are publicly available in the Zenodo database~\cite{zenodoDatabase}. 
%
%
\end{acknowledgments}
%
%
%
\appendix
\section{Calibration and error estimation}\label{sec:appA}
%

It is very important to highlight the difference between the phase mask for the calibration of the experiments, which we measure and characterize employing camera CCD2, and the phase information extracted from our experiments, which we obtain by controlling the incident beams onto the sample. 
When varying the phase difference between the two beams $\Delta \phi$ at a fixed angle, $F_{{\rm MS}}$ follows a sinusoidal curve, as illustrated in Fig.~\ref{fig:PhaseIter}. 
We extract the amplitude and phase of the control of $F_{{\rm MS}}$ for each angle $\gamma$. 
We do this by taking the Fourier transform of the raw data and filtering the point with frequency $f=1/2\pi$. 
This gives us the cosine-like behavior we expect from theory. 
The optimized amplitude of the filtered data is extracted as the amplitude of $F_{{\rm MS}}$. 
Conversely, the phase information of the mutual scattering is the phase difference at which $F_{{\rm MS}}$ is maximized.

We treat the error of the estimation of amplitude and phase separately. 
We consider the error range of our amplitude estimation to be the average difference between the filtered curve (solid black line at Fig.~\ref{fig:PhaseIter}) and the data (symbols at Fig.~\ref{fig:PhaseIter}), similar to the residual of a fitting process. 
Meanwhile, the error of the phase estimation is done through an iterative process. 
We do this by adding phase offsets to the experimental data and calculating the amplitude error for every new phase offset. 
The offset where the amplitude error is \textit{twice} (chosen based on experience) the original amplitude error, is chosen as the error range of our phase estimation. 

If we reach an offset equal to $2\pi$ and the amplitude error is still not doubled, we consider the phase to be \textit{undefined}, \textit{i.e.}, the data does not have a cosine-like shape, and we are not able to estimate the phase of the data. 

The most challenging step to extract the phase information is to calibrate the phase difference $\Delta \phi$ when changing the angle. 
This is because any change in path length generated by changing the angle will add a dephase. 
For example, given angle $\gamma_1$ with the set phase difference $\Delta \phi_{\gamma_1}$ at the modulator, changing to angle $\gamma_2$ without adapting the phase difference, any aberration, misalignment, or medium interface induces a dephase such that $\Delta \phi_{\gamma_2} \neq \Delta \phi_{\gamma_1}$ at the sample.

Calibrations of this induced dephase turn out to be tremendously difficult. 
We initially calibrate $\Delta \phi$ and $\gamma$ by measuring the interference fringes at CCD2. 
Besides technical limitations, such as noise, alignment, and camera resolution, we found that the camera cannot obtain the exact phase and angle information because the detector chip is covered by a protective glass~\cite{sony2009datasheet}, while the samples are embedded in PDMS.
These two media have different effects on the phase and angle of the incident light. 
Furthermore, to calibrate $\Delta \phi$ and $\gamma$ we need to estimate multiple parameters such as the depth of the sample in the medium, the orientation of the sample inside the medium, the roughness of the surface of the medium, any tilt of both cameras, any misalignment, etc. 
The estimation has so many degrees of freedom that it is not far from an educated guess, making it also more susceptible to confirmation bias. 

\begin{figure}[tbp]
    \centering
    \includegraphics[width=\columnwidth]{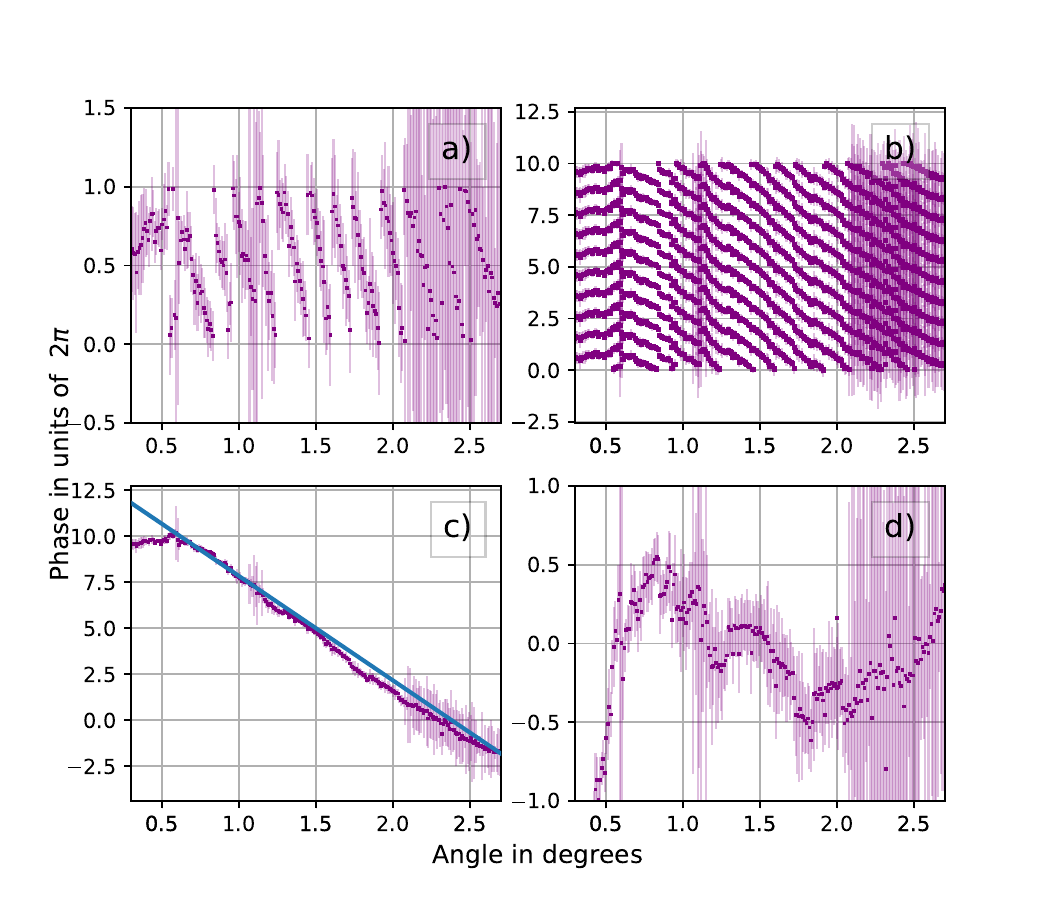}
    \caption{Steps for the calibration of the phase information of mutual scattering between angles. 
    a) We obtain the raw phase data from the phase dependency plots (Fig.~\ref{fig:PhaseIter}). 
    b) Then we add the same curve many times with jumps of $2\pi$. 
    c) From this curve, we select the correct $2\pi$ jumps to obtain a continuous line. 
    d) We finally do a linear regression (solid blue line) and subtract the linear regression to obtain the calibrated phase information. 
    }
    \label{fig:stepsPhase}
\end{figure}
%
There is a glimpse of hope when looking at the raw data of the phase information extracted when measuring mutual scattering. 
Fig.~\ref{fig:stepsPhase} illustrates the steps we currently follow to calibrate the phase information. 
Fig.~\ref{fig:stepsPhase}a shows the raw phase data for the dielectric sphere. 
Because of the periodicity of the phase, we know that $\phi = 2\pi+\phi$. 
Fig.~\ref{fig:stepsPhase}b shows the same raw data multiple times with jumps of $2\pi$. 
We select the $2\pi$ jump such that yields a continuous curve, shown in Fig.~\ref{fig:stepsPhase}c. 
We observe that there is a slow decaying envelope pattern in the phase, presumably coming from the dephases explained above, as it has a weak dependence on the angle (\textit{i.e.}, slow variations when changing $\gamma$) and larger than the phase information of the mutual scattering that we expect from theory for any sample. 
Thus, it is possible to reduce the effect of the envelope pattern by making a linear regression and subtracting it from the data in order to obtain the phase information shown in Fig.~\ref{fig:stepsPhase}d. 
This allows us to discuss the small variations in the data that have a strong dependence on the angle (\textit{i.e.}, fast variations when changing $\gamma$), which we attribute to the sample itself. 
We believe this correction is enough to get relevant information from the measured samples. 
We follow this procedure for all the samples present in this paper. 

%
\section{From total extinction to scattering amplitude}\label{sec:appB}
%

In our measurements, we extract the normalized mutual scattering component $F_{{\rm MS}}$ by measuring flux $F$, which is the integral of the current $J$ over the detector area $A$~\cite{lagendijk1996prep}. 
We have two incident waves with direction $\hat{k}_{{\rm in},1}$ and $\hat{k}_{{\rm in},2}$, and we integrate the current that goes in the direction $\hat{k}_{{\rm in},2}$. 
The mutual scattering in the direction of $\hat{k}_{{\rm in},2}$ is expressed as~\cite{Truong2023oe}
%
\begin{align}\label{eq:currentComponent1}
    J_{{\rm MS}}\hat{k}_{{\rm in},2}  =  A_1 A_2 {\rm Im} [ f(\hat{k}_{{\rm in},2},\hat{k}_{{\rm in},1}) e^{i(\Delta \phi_{2,1})} ],
\end{align}
%
where $f(\hat{k}_{{\rm in},i},\hat{k}_{{\rm in},j})$ is the scattering amplitude relating an incident wave at direction $\hat{k}_{{\rm in},j}$ and an outgoing wave at direction $\hat{k}_{{\rm in},i}$. 
The component $A_i$ is the amplitude of the wave at direction $\hat{k}_{{\rm in},i}$, and $\Delta \phi_{i,j}$ is the phase difference between the two incident waves.

The complex-valued scattering amplitude $f$ can be decomposed in the exponent form as $f=|f|\exp(i\phi_f)$. 
Thus, (\ref{eq:currentComponent1}) can be simplified as follows:
\begin{align}\label{eq:currentComponent2}
     J_{{\rm MS}}\hat{k}_{{\rm in},2}  =  A_1 A_2 |f_{2,1}| \sin ( \phi_{f} + \Delta \phi_{2,1}),
\end{align}
where we change the notation $f_{i,j}=f(\hat{k}_{{\rm in},i},\hat{k}_{{\rm in},j})$ for simplicity. 
The same notation follows for the phase component. 
We maximize this component if we set the phase difference as $\Delta \phi = \pi-\phi_f$. 
With this relation, we obtain $\phi_f$ in our experiment by scanning the phase and finding the maximum value. 
Furthermore, the maximum value of the current is $\max(J_{{\rm MS}}\hat{k}_{{\rm in},2})  =  A_1 A_2 |f|$. 
Consequently, we measure the amplitude of the incident waves $A_1, A_2$ separately, and hence we obtain $|f|$. 

For some cases, it is more informative to normalize the mutual scattering by the self-extinction $J_{{\rm self}}=A_{2}^2 \ {\rm Im} (f_{2,2})$, as we do for our experimental results. 
We define the normalized mutual scattering as 
\begin{align}\label{eq:currentComponent3}
     F_{\rm MS} = \frac{\int_{A} J_{{\rm MS}} \rm{d}S}{\int_{A} J_{{\rm in}} \rm{d}S},
\end{align}
where we integrate the current over the detector area $A$ to obtain the flux. 
Thus, the effect of the mutual scattering on the total extinction is clearer: if $1-F_{{\rm MS}}=2$, the total extinction is twice as large. 
If $1-F_{{\rm MS}}=0$, the total extinction vanishes. 
Experimentally, $F_{\rm MS}$ is obtained by measuring the flux for different scenarios, as shown in Eq.~\ref{eq:totalExtinction}.
%
%
%
%
\bibliographystyle{apsrev4-2}
\bibliography{References.bib}
\end{document}